\documentclass[aps,prd,twocolumn,superscriptaddress,nofootinbib,floatfix]{revtex4-2}
\usepackage{amsmath,amssymb,graphicx,bm}
\usepackage{xcolor}
\usepackage[colorlinks=true,linkcolor=blue,citecolor=blue,urlcolor=blue]{hyperref}

\begin{document}

\title{Electric-magnetic asymmetry of exact dyonic Euler-Heisenberg black holes: duality-blind shadows and magnetically controlled horizons}

\author{Vahideh Memari}
\affiliation{Cyprus International University, Nicosia, North Cyprus via Mersin 10, T\"urkiye}

\author{\.{I}zzet Sakall{\i}}
\email{izzet.sakalli@emu.edu.tr}
\affiliation{Physics Department, Eastern Mediterranean University, Famagusta 99628, North Cyprus via Mersin 10, T\"urkiye}

\begin{abstract}
Maxwell theory is invariant under electric-magnetic rotations, and the Reissner-Nordstr\"om black hole depends on its charges only through the sum of their squares. The Euler-Heisenberg vacuum-polarization terms break this symmetry, and an exact dyonic Einstein-Euler-Heisenberg black hole was found recently. We ask which properties remember the breaking. Outside the photon sphere, almost none do. The metric function is even under charge exchange through first order in the coupling, the first odd term appears at second order and falls off as the tenth inverse power of the radius, and the shifts it causes in the shadow, the Lyapunov exponent, and the innermost stable circular orbit stay at a few parts per million. The core behaves differently. Any magnetic charge drives the metric function to negative infinity at the center and so forces a horizon, while an electric configuration lighter than its field energy can be horizonless. Whether an extremal limit exists is decided by one function of the charges, the maximum of the derivative of the mass function, whose critical curve reaches the origin along the electric axis and meets the magnetic axis at a closed-form value. Two independent integrations confirm every number, and the shadow is tested against the Event Horizon Telescope image of Sagittarius A*.
\end{abstract}

\maketitle

\section{Introduction}\label{isec1}

The vacuum of quantum electrodynamics responds to strong fields as a nonlinear medium. Heisenberg and Euler summarized this response in an effective Lagrangian whose leading correction to Maxwell theory is quartic in the field strength \cite{Heisenberg:1936nmg,Schwinger:1951nm,Dunne:2004nc}. Coupled to general relativity, the same Lagrangian supplies a natural model for charged black holes whose electromagnetic sector is not linear, in the spirit of the older Born-Infeld construction \cite{Born:1934gh} and of the regular black holes sourced by nonlinear electrodynamics \cite{AyonBeato:1998ub,Bronnikov:2000vy}. Static solutions of the Einstein-Euler-Heisenberg (EEH) system were first studied by Yajima and Tamaki \cite{Yajima:2000kw} and later by Ruffini, Wu and Xue \cite{Ruffini:2013hia}. Their thermodynamics in anti-de Sitter space was worked out by Magos and Bret\'on \cite{Magos:2020ykt}, and the orbits of test particles around magnetically and dyonically charged EEH black holes were mapped by Magos Cort\'es, Bret\'on and Mac\'ias \cite{MagosCortes:2023tsv}. Geodesic studies in electromagnetic backgrounds have shown more generally that test bodies can tell electric from magnetic fields \cite{Halilsoy:2024plate,Halilsoy:2025chiral}, which makes the question of how a nonlinear black hole treats the two charges a natural one.

The magnetic sector has always been the easy one. With a monopole field the Bianchi identity fixes the field strength without solving any constraint, and the metric function follows in closed form. The electric sector is harder because the Gauss constraint of a quartic Lagrangian is a cubic equation for the field. A dyon combines the difficulties, since the two charges feed into the same cubic. For a long time the dyonic sector was known only numerically or through a large-radius expansion. The situation changed in 2026. Luo and collaborators obtained the purely electric EEH black hole in closed form and the dyonic one numerically \cite{Luo:2026ele}, an exact dyon was found on the special coupling line where the quadratic invariants combine into a perfect square \cite{Ahmed:2026ufj}, and finally the general dyonic solution was written in closed form with a Lauricella function \cite{Luo:2026dyo}. The last result covers the physical Euler-Heisenberg ratio of the two couplings.

These developments put a sharp question within reach. Maxwell electrodynamics is invariant under duality rotations that mix electric and magnetic fields, and the Reissner-Nordstr\"om black hole depends on its charges only through $Q^2+P^2$. Nonlinear theories that keep this invariance are rare and must satisfy the Gaillard-Zumino-Gibbons-Rasheed condition \cite{Gaillard:1981rj,Gibbons:1995cv}; the conformal ModMax theory is the best known recent example \cite{Bandos:2020jsw}. The Euler-Heisenberg Lagrangian does not satisfy that condition, so its black holes need not treat the two charges alike. How strongly, and where, does the asymmetry show up?

The existing phenomenology cannot answer this. Studies of charged and dyonic EEH black holes have worked with the metric function at first order in the coupling or with the exact magnetic one \cite{MagosCortes:2023tsv,Kala:2025qnm,Myung:2025eeh}, since no exact dyonic form was available, and the optical and thermodynamic analyses of the charged EEH family \cite{Allahyari:2019jqz,Jafarzade:2025eeh,Theodosopoulos:2024eh,Mustafa:2025eh} rest on the same truncation or on the purely magnetic case, where the solution is exact. A truncation at first order depends on the charges only through $Q^2+P^2$ and through a single quartic combination that is itself symmetric under $Q\leftrightarrow P$, as we show below. It is therefore blind to duality breaking by construction. The exact solution of Ref.~\cite{Luo:2026dyo}, in turn, has been written down but not yet probed for the observables that separate the two charges.

The answer we find has two parts that point in opposite directions. Everything an observer at large radius can measure is nearly duality-blind. The odd part of the metric function enters only at second order in the coupling and only as $r^{-10}$, and we derive its coefficient in closed form and confirm it numerically. The shifts it produces in the shadow, the Lyapunov exponent and the innermost stable circular orbit (ISCO) are at most of order $10^{-6}$ in units of $M$ for the couplings considered. The core, by contrast, is controlled by the magnetic charge alone. Any $P\neq0$ drives the metric function to $-\infty$ at the center, independently of the electric charge and of the second coupling, so a dyon always has an odd number of horizons and can never be a naked singularity, whereas an electric configuration lighter than its own field energy has an even number of horizons and can be horizonless. The existence of an extremal limit, and with it the three-horizon window and the jump of the event horizon at the edge of that window, is decided by one function of the charges. We give its critical value in closed form on the magnetic axis.

The gap this paper closes is therefore specific: the leading-order treatments of dyonic EEH black holes cannot register duality breaking, and the exact solution has not yet been used to locate where the breaking lives and where it is screened from an observer at large radius.

The paper is organized as follows. Section~\ref{isec2} fixes conventions, recasts the exact solution in a form suited to computation, and derives its large-radius expansion with the duality-odd term. Section~\ref{isec3} studies the core, the horizon count and the extremality criterion. Section~\ref{isec4} treats the thermodynamics, including the first law and the Smarr relation with the coupling as a variable. Section~\ref{isec5} covers the geodesic structure: circular orbits, precessing and zoom-whirl bound orbits, and radial infall toward the core. Section~\ref{isec6} treats lensing, the shadow and ray tracing beyond the charged branch. Section~\ref{isec8} summarizes the results together with the consistency checks and the scope of the analysis. We use $G=c=1$ and the signature $(-,+,+,+)$.

\section{Field equations and the exact dyonic solution}\label{isec2}

This section sets up the model, writes the exact solution in two independent representations, and extracts the large-radius structure that controls every exterior observable. The aim is to isolate, already at the level of the metric function, which pieces are even and which are odd under the exchange of the charges.

\subsection{Action, ansatz and the Gauss constraint}\label{isec2a}

We take the action \cite{Luo:2026dyo,Ahmed:2026ufj}
\begin{equation}
S=\frac{1}{16\pi}\int d^4x\,\sqrt{-g}\,\big[R-\mathcal{L}(\mathcal{F},\mathcal{G})\big],\label{eq:action}
\end{equation}
with
\begin{equation}
\mathcal{L}=\mathcal{F}-a\mathcal{F}^2-b\,\mathcal{G}^2,\qquad \mathcal{F}=F_{\mu\nu}F^{\mu\nu},\qquad \mathcal{G}=\tilde F_{\mu\nu}F^{\mu\nu}.\label{eq:lag}
\end{equation}
Here $\tilde F^{\mu\nu}=\frac12\eta^{\mu\nu\rho\sigma}F_{\rho\sigma}$, and $a$ and $b$ have dimensions of length squared. The Heisenberg-Euler weights of $(\mathbf{E}^2-\mathbf{B}^2)^2$ and $(\mathbf{E}\cdot\mathbf{B})^2$ fix $b=7a/4$ \cite{Heisenberg:1936nmg,Dunne:2004nc}; we keep $b$ general where it costs nothing and specialize at the end. Metric variation gives
\begin{equation}
G_{\mu\nu}=2\big(\mathcal{L}_{\mathcal{F}}F_{\mu\lambda}F_{\nu}{}^{\lambda}+\mathcal{L}_{\mathcal{G}}F_{\mu\lambda}\tilde F_{\nu}{}^{\lambda}\big)-\tfrac12 g_{\mu\nu}\mathcal{L},\label{eq:einstein}
\end{equation}
and variation of the potential gives
\begin{equation}
\nabla_\mu\big(\mathcal{L}_{\mathcal{F}}F^{\mu\nu}+\mathcal{L}_{\mathcal{G}}\tilde F^{\mu\nu}\big)=0 .\label{eq:maxwell}
\end{equation}
The bulk equations do not depend on the Gibbons-Hawking-York term, which we would add for a Euclidean treatment.

For the static, spherically symmetric sector we write
\begin{equation}
ds^2=-f(r)\,dt^2+\frac{dr^2}{f(r)}+r^2d\Omega^2,\qquad f(r)=1-\frac{2m(r)}{r},\label{eq:metric}
\end{equation}
and $A=V(r)\,dt+P\cos\theta\,d\phi$. The choice $g_{tt}g_{rr}=-1$ is exact here because the radial electric field $E=-V'$ and the radial magnetic field $B=P/r^2$ give $T^t{}_t=T^r{}_r$. The Bianchi identity holds automatically, so $B$ needs no further work. Equation~\eqref{eq:maxwell} integrates once to the Gauss constraint
\begin{equation}
E+4aE^3+4cB^2E=\frac{Q}{r^2},\qquad c\equiv 2b-a ,\label{eq:gauss}
\end{equation}
with $Q$ the electric charge. For $a>0$ and $c\geq0$ the cubic has a single real root,
\begin{equation}
E(r)=2\sqrt{\frac{p}{3}}\,\sinh\!\Big[\frac13\,\mathrm{arsinh}\Big(\frac{3\sqrt3\,Q}{8a\,r^2p^{3/2}}\Big)\Big],\label{eq:Eroot}
\end{equation}
with
\begin{equation}
p=\frac{1}{4a}\Big(1+\frac{4cP^2}{r^4}\Big).\label{eq:p}
\end{equation}
This form is free of the cancellation that the two-cube-root Cardano expression suffers at large $r$. For EEH, $c=5a/2$.

Equation~\eqref{eq:gauss} is where the two charges part ways. The magnetic field is fixed by topology, while the electric field must be solved for and is screened by both nonlinear terms. It is convenient to read the theory through its constitutive relations. With $D=Q/r^2$ and the reduced Lagrangian
\begin{equation}
\ell(E,B)=\tfrac12\big[E^2-B^2+2a(E^4+B^4)+4cE^2B^2\big],\label{eq:ell}
\end{equation}
one has $D=\partial\ell/\partial E$, which reproduces Eq.~\eqref{eq:gauss}, and the magnetic intensity
\begin{equation}
H=-\frac{\partial\ell}{\partial B}=B\big(1-4aB^2-4cE^2\big).\label{eq:H}
\end{equation}

\subsection{Mass function in two representations}\label{isec2b}

The $tt$ component of Eq.~\eqref{eq:einstein} gives $m'(r)=r^2u$, where $u=ED-\ell$ is the energy density. Using Eq.~\eqref{eq:gauss} and the variable $z\equiv r^2E$, this becomes
\begin{equation}
m'(r)=\frac{P^2+Qz}{2r^2}+\frac{a\,(z^4-P^4)}{r^6},\label{eq:mprime}
\end{equation}
in agreement with Ref.~\cite{Luo:2026dyo}. Fixing the ADM mass $M$ gives the first representation,
\begin{equation}
m(r)=M-\int_r^\infty m'(s)\,ds .\label{eq:mR}
\end{equation}
Equations~\eqref{eq:Eroot}, \eqref{eq:mprime} and \eqref{eq:mR} define route I: a one-dimensional quadrature of an elementary integrand.

The second representation trades $r$ for $z$. Equation~\eqref{eq:gauss} inverts to
\begin{equation}
r^4=\frac{4z\,h(z)}{Q-z},\qquad h(z)=az^2+cP^2,\qquad 0<z<Q,\label{eq:rz}
\end{equation}
and the mass function obeys \cite{Luo:2026dyo}
\begin{equation}
\frac{dm}{dz}=\frac{\mathcal{P}_8(z)}{16\sqrt2\,z^{9/4}h(z)^{9/4}(Q-z)^{3/4}},\label{eq:dmdz}
\end{equation}
where $\mathcal{P}_8$ is the eighth-order polynomial tabulated there. Its antiderivative is a sum of Lauricella functions $F_D^{(3)}$. Since $F_D^{(3)}$ is not available in standard computer algebra systems and its series diverges once $z^2>cP^2/a$, we evaluate the equivalent Euler integral directly. The substitution $z=Q-w^4$ removes the endpoint singularity:
\begin{equation}
m(z)=M-\int_0^{(Q-z)^{1/4}}\frac{4\,\mathcal{P}_8(Q-w^4)\,dw}{16\sqrt2\,(Q-w^4)^{9/4}\,h(Q-w^4)^{9/4}} .\label{eq:mZ}
\end{equation}
This is route II. It never solves the cubic and it integrates in a different variable, so its agreement with route I tests both the closed-form polynomial and the quadrature. We evaluate it at 30-digit precision. The two routes agree to $10^{-14}$ in $f(r)$ at every radius we sampled, from $r=0.4M$ to $40M$. A third, independent evaluation in Wolfram Language reproduces the horizon, photon-sphere and shadow radii of Table~\ref{tab2} to all printed digits.

Before any new result we checked the background against every limit available. Table~\ref{tab1} lists these anchors. The Schwarzschild values $b_c=3\sqrt3M$, $\lambda=1/(3\sqrt3M)$ and $r_{\rm ISCO}=6M$ with $\mathcal{E}=\sqrt{8/9}$ and $\mathcal{L}=2\sqrt3M$ come out to the precision shown, and so do the Reissner-Nordstr\"om limit, the exact magnetic metric derived below, the integrable locus $b=a/2$ of Ref.~\cite{Ahmed:2026ufj}, and the closed forms of Sec.~\ref{isec3}.

\begin{table}[t]
\caption{Anchors reproduced before any new result. Route I is the radial quadrature, Eq.~\eqref{eq:mR}; route II is the $z$ representation, Eq.~\eqref{eq:mZ}; closed forms are as cited.}
\label{tab1}
\begin{ruledtabular}
\footnotesize
\setlength{\tabcolsep}{3pt}
\begin{tabular}{lll}
\textbf{Check} & \textbf{Expected} & \textbf{Obtained}\\
\hline
Schwarzschild $b_c$ & $3\sqrt3$ & $5.19615242$\\
Schwarzschild $\lambda$ & $1/(3\sqrt3)$ & $0.19245009$\\
Schwarzschild $r_{\rm ISCO}$ & $6$ & $6.0000000$\\
RN $r_{\rm h}$ ($a\to0$, $Q=P=0.4$) & $1.824621$ & $1.824621$\\
$r_{\rm h}$ from Eq.~\eqref{eq:fmag}, $P=0.4$ & $1.91653584$ & $1.91653584$\\
Locus $b=a/2$, $m(1)$ & $0.8306744494$ & $0.8306744494$\\
$M_{\rm em}$, Eq.~\eqref{eq:Mem} & $3.15625602334$ & $3.15625602334$\\
$\mathcal{M}(0,P_c)$ & $0.5$ & $0.5000000$\\
Route I vs II in $f(r)$ & $0$ & $\leq10^{-14}$\\
Eq.~\eqref{eq:Deltaf}, $r=40$ & $1$ & $0.9999998$\\
Smarr, Eq.~\eqref{eq:smarr} & $0.9935450972$ & $0.9935450972$\\
\end{tabular}
\end{ruledtabular}
\end{table}

\subsection{Large-radius expansion and the duality-odd term}\label{isec2c}

Iterating Eq.~\eqref{eq:gauss} in powers of $1/r$ and inserting the result into Eq.~\eqref{eq:mprime} gives
\begin{align}
m'(r)={}&\frac{Q^2+P^2}{2r^2}-\frac{\mathcal{K}}{r^6}+\frac{8Q^2X^2}{r^{10}}\nonumber\\
&-\frac{32Q^2X^2(X+2aQ^2)}{r^{14}}+\mathcal{O}(r^{-18}),\label{eq:mpseries}
\end{align}
where
\begin{align}
\mathcal{K}&=aQ^4+2cQ^2P^2+aP^4\nonumber\\
&=a(Q^2-P^2)^2+4bQ^2P^2,\label{eq:K}\\
X&=aQ^2+cP^2 .\label{eq:X}
\end{align}
Integrating,
\begin{align}
f(r)={}&1-\frac{2M}{r}+\frac{Q^2+P^2}{r^2}-\frac{2\mathcal{K}}{5r^6}\nonumber\\
&+\frac{16Q^2X^2}{9r^{10}}-\frac{64Q^2X^2(X+2aQ^2)}{13\,r^{14}}+\cdots .\label{eq:fseries}
\end{align}
For EEH, $\mathcal{K}=a(Q^4+5Q^2P^2+P^4)$. The first three terms are the Reissner-Nordstr\"om metric function, and the $r^{-6}$ term is the familiar leading Euler-Heisenberg correction \cite{Yajima:2000kw,Ruffini:2013hia}. The expansion carries two structural facts that the rest of the paper builds on.

First, $\mathcal{K}$ is symmetric under $Q\leftrightarrow P$ for any $a$ and $b$. The entire first-order metric is duality-even, even though the Lagrangian is not duality invariant. An analysis built on the first-order metric therefore cannot register the breaking.

Second, every term beyond $r^{-6}$ carries a factor $Q^2$. The magnetic tower terminates: for $Q=0$ the quadrature closes exactly and
\begin{equation}
f_{Q=0}(r)=1-\frac{2M}{r}+\frac{P^2}{r^2}-\frac{2aP^4}{5r^6},\label{eq:fmag}
\end{equation}
whereas the electric tower does not terminate. The asymmetry comes from the screening of $E$ in Eq.~\eqref{eq:gauss}; $B$ is never screened. The leading odd part follows from Eq.~\eqref{eq:fseries},
\begin{align}
f(Q,P)-f(P,Q)={}&\frac{16}{9\,r^{10}}\Big[Q^2(aQ^2+cP^2)^2\nonumber\\
&-P^2(aP^2+cQ^2)^2\Big]+\mathcal{O}(r^{-14}),\label{eq:Deltaf}
\end{align}
which for EEH reads
\begin{align}
f(Q,P)-f(P,Q)={}&\frac{4a^2}{9r^{10}}\,(Q^2-P^2)(4Q^4-Q^2P^2+4P^4)\nonumber\\
&+\mathcal{O}(a^3r^{-14}).\label{eq:DeltafEEH}
\end{align}
We checked Eq.~\eqref{eq:Deltaf} against route II for $(Q,P)=(0.7,0.3)M$ and $a=0.1M^2$: the ratio of the numerical difference to the formula is $0.99994$, $0.999996$ and $0.9999998$ at $r=10M$, $20M$ and $40M$, approaching unity as $1-\mathcal{O}(r^{-4})$, as the next term requires.

\begin{figure*}[t]
\centering
\includegraphics[width=\textwidth]{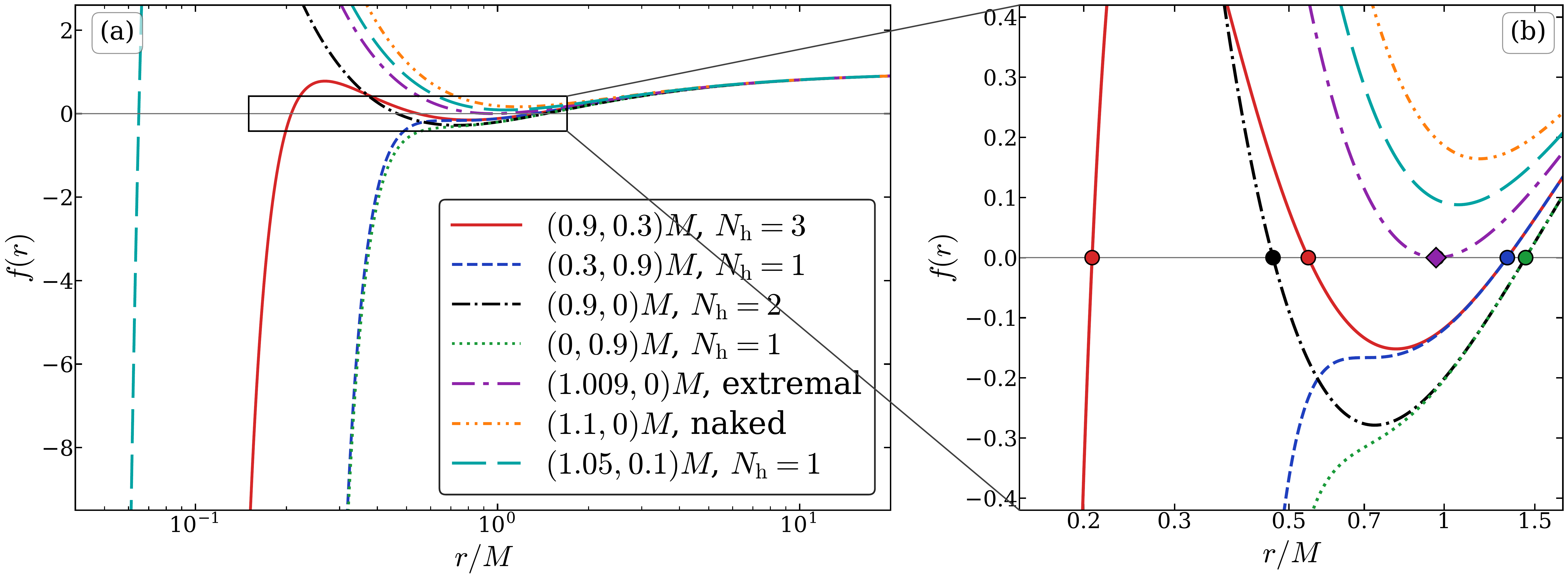}
\caption{Metric function for $a=0.05M^2$ and $b=7a/4$ at $M=1$, covering every horizon structure the family admits: three horizons for the electrically dominated dyon $(0.9,0.3)M$, a single horizon for its dual $(0.3,0.9)M$ and for the pure magnetic case $(0,0.9)M$, two horizons for the pure electric case $(0.9,0)M$, the extremal electric configuration $(1.009,0)M$, the horizonless electric configuration $(1.1,0)M$, and the dyon $(1.05,0.1)M$ beyond the charged branch, which keeps a single small horizon at $r=0.065M$. (a) Full range on a logarithmic radial axis; every curve with $P\neq0$ ends at $-\infty$ and every curve with $P=0$ at $+\infty$. (b) Magnification of the boxed region. Circles mark the simple zeros of $f$ and the diamond the degenerate zero of the extremal case at $r=0.966M$. The outer horizons of $(0.9,0.3)M$ and $(0.3,0.9)M$ coincide at $r=1.326M$, so their markers overlap.}
\label{fig1}
\end{figure*}

Equation~\eqref{eq:Deltaf} is the central equation of the exterior analysis, and its two features, second order in the coupling and tenth inverse power of $r$, are universal across the quadratic family~\eqref{eq:lag}. Only the numerical coefficient depends on $b/a$. At the locus $b=a/2$ of Ref.~\cite{Ahmed:2026ufj} it reduces to $16a^2(Q^6-P^6)/(9r^{10})$, which is still nonzero, so that solution breaks duality as well. Figure~\ref{fig1} displays the metric function for each horizon structure the family admits at fixed mass. Outside the outer horizons a dyon and its dual are indistinguishable at plotting accuracy, while inside every curve with $P\neq0$ turns down to $-\infty$ and every purely electric curve turns up to $+\infty$, which anticipates the core analysis of the next section.

\section{Core structure, horizon count and the extremality criterion}\label{isec3}

The duality-odd part of the metric function vanishes quickly at large radius, but nothing protects it near the center. This section shows that the inner geometry is set by the magnetic charge alone and derives a criterion for the existence of an extremal limit that depends only on the charges.

\subsection{Central asymptotics}\label{isec3a}

For $P\neq0$ and $c>0$, Eq.~\eqref{eq:rz} gives $z\simeq Qr^4/(4cP^2)$ as $r\to0$, so $E\to0$ at the center. Equation~\eqref{eq:mprime} then reduces to $m'\simeq -aP^4/r^6+P^2/(2r^2)$, and
\begin{equation}
m(r)\simeq \frac{aP^4}{5r^5}-\frac{P^2}{2r}+m_0,\qquad f(r)\simeq-\frac{2aP^4}{5r^6}\quad (r\to0).\label{eq:core}
\end{equation}
The leading term is independent of $Q$ and of $b$. It survives at $c=0$, where $E$ is no longer suppressed at the center but grows only as $r^{-2/3}$, too slowly to compete with the $aP^4/r^6$ term.

For $P=0$ the Gauss constraint gives $E\simeq[Q/(4ar^2)]^{1/3}$ near the center and $m'\simeq\frac34 Q(Q/4a)^{1/3}r^{-2/3}$, which is integrable. The mass function tends to the finite value $m(0)=M-M_{\rm em}$, where the electrostatic self-energy follows from the $z$ representation as a sum of Beta functions,
\begin{align}
M_{\rm em}&=\frac{9B(\frac14,\frac14)-9B(\frac54,\frac14)+2B(\frac94,\frac14)}{16\sqrt2}\,\frac{Q^{3/2}}{a^{1/4}}\nonumber\\
&=\frac{\Gamma(\frac14)^2}{3\sqrt{2\pi}}\,\frac{Q^{3/2}}{a^{1/4}} .\label{eq:Mem}
\end{align}
The prefactor is $1.74803837$. For $Q=0.9M$ and $a=0.05M^2$ Eq.~\eqref{eq:Mem} gives $3.156256023341M$, while route I gives $3.156256023344M$.

\subsection{Horizon parity}\label{isec3b}

Since $f$ is continuous on $(0,\infty)$ and $f(\infty)=1$, the number of simple zeros is fixed in parity by the sign of $f(0^+)$. Equation~\eqref{eq:core} gives the first statement below, and $f\simeq 1-2(M-M_{\rm em})/r$ for $P=0$ gives the second:
\begin{equation}
N_{\rm h}\ \text{is}\ \begin{cases}\text{odd},\ N_{\rm h}\geq1, & P\neq0,\\ \text{odd}, & P=0,\ M>M_{\rm em},\\ \text{even, possibly zero}, & P=0,\ M<M_{\rm em}.\end{cases}\label{eq:parity}
\end{equation}
In the quadratic model a dyon is never a naked singularity, whatever its charge-to-mass ratio, so the magnetic sector enforces the cosmic-censorship expectation \cite{Penrose:1969pc} on its own, while an electric configuration lighter than its own field energy can be one. The condition $M<M_{\rm em}$ is necessary but not sufficient for this. At $a=0.05M^2$ the pure electric configurations $(0.9,0)M$ and $(1.1,0)M$ both lie below their field energy, with $M_{\rm em}=3.156M$ and $4.265M$, yet the first has two horizons and the second none (Fig.~\ref{fig1}); the pair merges at the extremal charge $Q=1.009M$. Here the asymmetry is of order one. Swapping the charges of a horizonless electric configuration produces a black hole.

The transition is continuous in $P$. Adding a small magnetic charge to a horizonless electric configuration creates a horizon whose radius follows from balancing the two leading terms of $f$ near the center,
\begin{equation}
r_1\simeq\Big[\frac{aP^4}{5(M_{\rm em}-M)}\Big]^{1/5},\qquad P\to0 .\label{eq:r1}
\end{equation}
For $Q=1.2M$ and $a=0.05M^2$, where $M_{\rm em}=4.859M$ and the electric configuration has no horizon, the numerical $r_1$ approaches Eq.~\eqref{eq:r1} as $P$ decreases from $10^{-1}M$ to $10^{-5}M$, with a ratio falling from $1.235$ to $1.015$. Fig.~\ref{fig3}(a) shows the comparison.

\subsection{The extremality functional}\label{isec3c}

At fixed charges the horizon condition $m(r_{\rm h})=r_{\rm h}/2$ defines the mass as a function of the horizon radius,
\begin{equation}
M(r_{\rm h})=\frac{r_{\rm h}}{2}+\int_{r_{\rm h}}^{\infty}m'(s)\,ds,\qquad \frac{dM}{dr_{\rm h}}=\frac12-m'(r_{\rm h}) .\label{eq:Mrh}
\end{equation}
A degenerate horizon requires $m'(r_{\rm h})=1/2$. This motivates the extremality functional
\begin{equation}
\mathcal{M}(Q,P)\equiv\sup_{r>0}\,m'(r),\label{eq:calM}
\end{equation}
which depends on the charges and couplings but not on $M$. If $\mathcal{M}<1/2$, $M(r_{\rm h})$ is monotonic and every mass has exactly one horizon with $f'(r_{\rm h})>0$; the family has no extremal member. If $\mathcal{M}>1/2$ and $P\neq0$, the equation $m'(r)=1/2$ has at least two roots, since $m'\to-\infty$ at the center and $m'\to0$ at infinity; in every case we scanned it has exactly two, $r_-<r_+$. Then $M(r_{\rm h})$ has a local maximum $M_{\max}=M(r_-)$ and a local minimum $M_{\min}=M(r_+)$, and three horizons exist for $M_{\min}<M<M_{\max}$.

On the magnetic axis Eq.~\eqref{eq:fmag} gives $m'=P^2/(2r^2)-aP^4/r^6$, with maximum $P/(3\sqrt{6a})$ at $r^4=6aP^2$. The critical charge, radius and mass are
\begin{equation}
P_c=3\sqrt{\tfrac{3a}{2}},\qquad r_c=3\sqrt a,\qquad M_c=\tfrac{18}{5}\sqrt a ,\label{eq:Pc}
\end{equation}
at which the three horizons merge into one of third order. Extremal configurations of this kind matter beyond the static family, since extremal Reissner-Nordstr\"om geometries can arise from colliding electromagnetic waves \cite{Memari:2026extremal}. On the electric axis $m'\propto r^{-2/3}$ near the center, so $\mathcal{M}=\infty$ for every $Q>0$ and an extremal limit always exists, as for Reissner-Nordstr\"om.

\begin{figure*}[t]
\centering
\includegraphics[width=\textwidth]{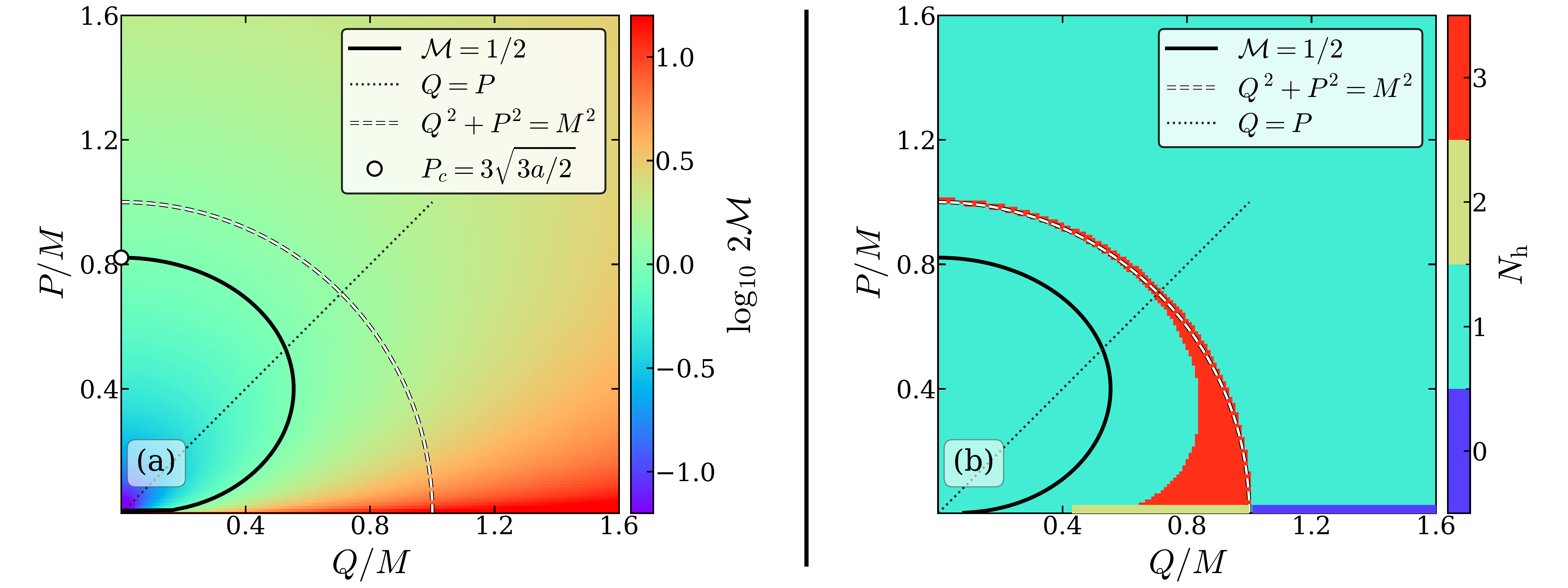}
\caption{(a) Extremality functional $\mathcal{M}=\sup_r m'(r)$ over the charge plane for $a=0.05M^2$, $b=7a/4$, shown as $\log_{10}2\mathcal{M}$. The solid black contour marks $\mathcal{M}=1/2$; inside it no member of the family has a degenerate horizon. It meets the magnetic axis at $P_c=3\sqrt{3a/2}=0.8216M$ (open circle) and approaches the origin along the electric axis. The white dashed arc is the Reissner-Nordstr\"om extremality circle and the dotted line the diagonal $Q=P$. (b) Number of horizons at fixed $M$ for the same couplings, with the same color scale family. The three-horizon band hugs the extremality circle and is much wider on the electric side; the strip along the electric axis shows the even counts, two horizons for $0.418M<Q<1.009M$ and none beyond. The critical curve of panel (a) is repeated in black.}
\label{fig2}
\end{figure*}

Rescaling $r\to\sqrt a\,r$ shows that $\mathcal{M}$ depends on the charges only through $Q/\sqrt a$ and $P/\sqrt a$ at fixed $b/a$. The critical curve $\Gamma_c$, defined by $\mathcal{M}=1/2$, is therefore a single universal curve in these units. It runs from the origin, approached along the electric axis, to $P_c$ on the magnetic axis. Figure~\ref{fig2}(a) shows $\mathcal{M}$ over the charge plane for $a=0.05M^2$, where $P_c=0.82158M$. The figure is strongly asymmetric about the diagonal: a configuration with $(Q,P)=(0.9,0.3)M$ has $\mathcal{M}=1.014$, its dual $(0.3,0.9)M$ has $\mathcal{M}=0.582$, and the region without an extremal limit lies entirely on the magnetic side of the curve.

\subsection{Horizon count at fixed mass and the jump of the event horizon}\label{isec3d}

Figure~\ref{fig2}(b) shows the number of horizons at fixed $M$. The three-horizon band follows the extremality circle $Q^2+P^2\simeq M^2$ of the Reissner-Nordstr\"om solution, but its width is not duality symmetric. At $\sqrt{Q^2+P^2}=0.95M$ the window $M_{\max}-M_{\min}$ is $0.485$ for $\arctan(P/Q)=0.12$ and $0.012$ on the magnetic axis. Even counts appear only on the electric axis, as Eq.~\eqref{eq:parity} requires, and they switch on where $M=M_{\rm em}$, that is at $Q_*=[M a^{1/4}\,3\sqrt{2\pi}/\Gamma(\tfrac14)^2]^{2/3}=0.418M$ for $a=0.05M^2$. Below $Q_*$ the electric black hole has a single horizon like Schwarzschild; above it there are two horizons up to the extremal value $Q=1.009M$ and none beyond. The band along the bottom edge of Fig.~\ref{fig2}(b) shows these three segments.

\begin{figure*}[t]
\centering
\includegraphics[width=\textwidth]{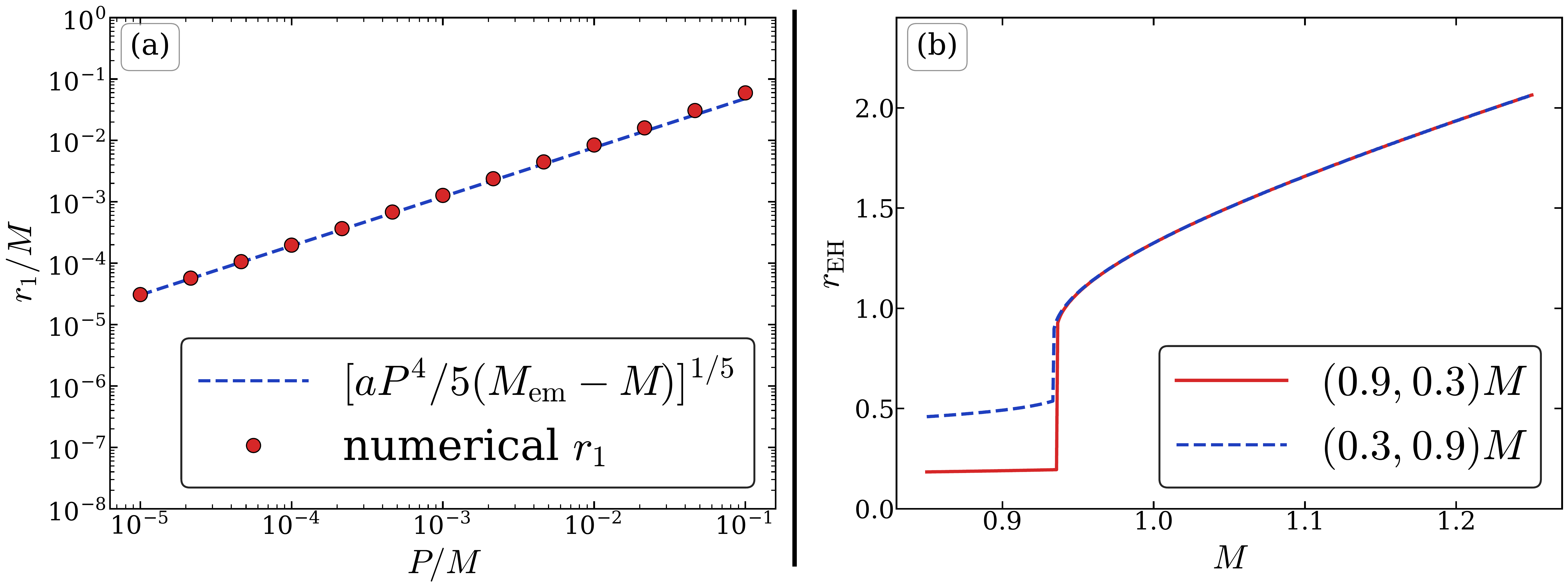}
\caption{(a) Radius of the horizon created by a small magnetic charge in an otherwise horizonless electric configuration, $Q=1.2M$, $a=0.05M^2$, compared with Eq.~\eqref{eq:r1}. (b) Event-horizon radius against the mass at fixed charges $(Q,P)=(0.9,0.3)$ and $(0.3,0.9)$, with $a=0.05$ in the same length unit. Above the three-horizon window the two curves coincide; below it the event horizon jumps to the inner, magnetically supported horizon, whose radius differs by a factor of almost three between the two orientations.}
\label{fig3}
\end{figure*}

Leaving the band toward lower mass does not create a naked singularity. The two outer horizons merge and disappear, and the event horizon jumps inward to the magnetically supported horizon. For $(Q,P)=(0.9,0.3)$ the merger happens at $M_{\min}=0.93583$, where the event horizon jumps from $r=0.88573$ to $r=0.19512$; for the dual $(0.3,0.9)$ it happens at $M_{\min}=0.93364$, with a jump from $0.86179$ to $0.53827$. The radii are in the length unit that fixes the charges. Figure~\ref{fig3}(b) shows both curves. The jump occurs at a zero-temperature point of the large branch, so the third law prevents an evaporating black hole from crossing it; it is a feature of the family, not of a single history.

\section{Thermodynamics}\label{isec4}

The extremality functional has a direct thermodynamic reading because $dM/dr_{\rm h}$ is proportional to the temperature. This section writes the first law and the Smarr relation with closed-form potentials, verifies both numerically, and follows the branch structure that the core produces.

\subsection{Temperature, first law and Smarr relation}\label{isec4a}

The Hawking temperature \cite{Hawking:1975vcx} is
\begin{align}
T&=\frac{f'(r_{\rm h})}{4\pi}=\frac{1-2m'(r_{\rm h})}{4\pi r_{\rm h}}\nonumber\\
&=\frac{1}{4\pi r_{\rm h}}\Big[1-\frac{P^2+Qz_{\rm h}}{r_{\rm h}^2}-\frac{2a(z_{\rm h}^4-P^4)}{r_{\rm h}^6}\Big],\label{eq:T}
\end{align}
with $z_{\rm h}=r_{\rm h}^2E(r_{\rm h})$. It is elementary even though $M$ itself requires the Lauricella function. Since the gravitational action is Einstein-Hilbert, the entropy is $S=\pi r_{\rm h}^2$ \cite{Bekenstein:1973ur}, and Eq.~\eqref{eq:Mrh} gives $dM=T\,dS$ at fixed charges. Treating the charges and the coupling as variables, with $b/a=7/4$ held fixed, the first law reads \cite{Rasheed:1997ns,Bokulic:2021dtz,Gunasekaran:2012dq}
\begin{equation}
dM=T\,dS+\Phi\,dQ+\Psi\,dP+\mathcal{A}\,da ,\label{eq:firstlaw}
\end{equation}
and differentiating $m'=r^2u$ under the integral of Eq.~\eqref{eq:Mrh} gives the potentials in closed integral form,
\begin{align}
\Phi&=\int_{r_{\rm h}}^{\infty}E\,dr,\qquad \Psi=\int_{r_{\rm h}}^{\infty}H\,dr,\label{eq:PhiPsi}\\
\mathcal{A}&=-\int_{r_{\rm h}}^{\infty}r^2\big(E^4+5E^2B^2+B^4\big)\,dr,\label{eq:calA}
\end{align}
with $H$ from Eq.~\eqref{eq:H}. In Maxwell theory $\Psi=P/r_{\rm h}=\Phi(Q\to P)$. Here the integrands differ, since $E$ is screened by Eq.~\eqref{eq:gauss} and $H$ by Eq.~\eqref{eq:H}. Under $r\to\lambda r$ the variables scale as $S\to\lambda^2S$, $Q,P\to\lambda Q,\lambda P$ and $a\to\lambda^2a$, and Euler's theorem gives the Smarr relation \cite{Kastor:2009wy,Gunasekaran:2012dq}
\begin{equation}
M=2TS+\Phi Q+\Psi P+2a\mathcal{A}.\label{eq:smarr}
\end{equation}
For $r_{\rm h}=1.3$, $(Q,P)=(0.9,0.3)$ and $a=0.05$, finite differences of $M(r_{\rm h},Q,P,a)$ give $\partial_QM=0.683160120$, $\partial_PM=0.224393231$ and $\partial_aM=-0.049281171$, and the integrals~\eqref{eq:PhiPsi} and \eqref{eq:calA} return the same digits. The right side of Eq.~\eqref{eq:smarr} equals $M=0.9935450972$ to all ten printed digits. The dual point gives $\Phi=0.224478$ and $\Psi=0.682582$, close to the swapped values but not equal.

\subsection{Branches, heat capacity and free energy}\label{isec4b}

\begin{figure*}[t]
\centering
\includegraphics[width=\textwidth]{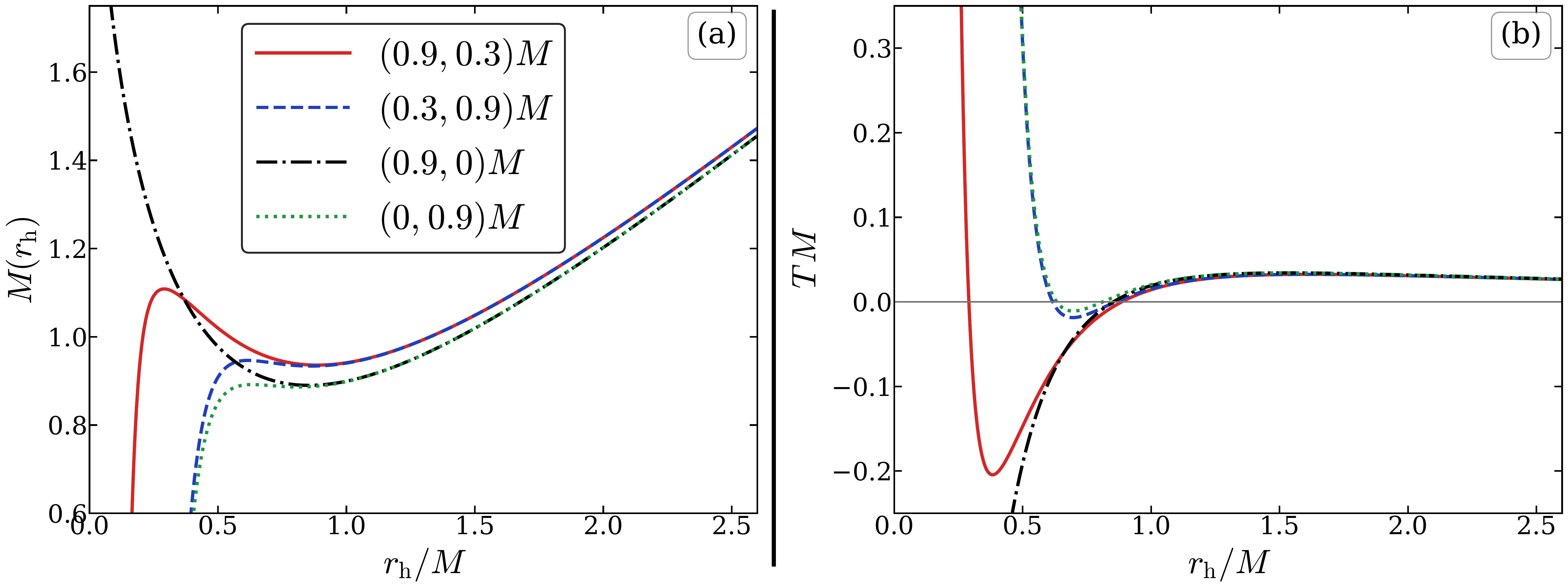}
\caption{(a) Mass as a function of the horizon radius at fixed charges, $a=0.05M^2$. Extrema mark degenerate horizons. The curve for $(0.9,0.3)M$ has a tall maximum near $r_{\rm h}=0.29M$, its dual a shallow one near $0.62M$, and the pure electric curve none, tending instead to $M_{\rm em}$. (b) Hawking temperature, Eq.~\eqref{eq:T}. The negative pieces are inner horizons. The small-black-hole branch at left exists only when $P\neq0$, and all four curves merge on the large branch.}
\label{fig4}
\end{figure*}

Figure~\ref{fig4} plots $M(r_{\rm h})$ and $T(r_{\rm h})$ for the dual pair and for the two single-charge configurations with the same magnitude. Three features stand out. First, for $P\neq0$ the curve $M(r_{\rm h})$ falls to $-\infty$ as $r_{\rm h}\to0$, which is the thermodynamic form of Eq.~\eqref{eq:parity}, and there is a small-black-hole branch with
\begin{equation}
T\simeq\frac{aP^4}{2\pi r_{\rm h}^7},\qquad C_{Q,P}=2\pi r_{\rm h}T\Big(\frac{\partial T}{\partial r_{\rm h}}\Big)^{-1}<0,\qquad r_{\rm h}\to0 .\label{eq:smallbranch}
\end{equation}
The electric configuration has no such branch, since its $M(r_{\rm h})$ tends to $M_{\rm em}$. Second, the middle branch with $T<0$ consists of inner horizons. Third, the large branch is the familiar charged one: a near-extremal piece with $C>0$ ends at a Davies point \cite{Davies:1977bgr}, beyond which $C<0$.

The large branch is nearly duality-blind. Its Davies point sits at $r_{\rm h}=1.6121$, $T=0.032544$ for $(0.9,0.3)$ and at $r_{\rm h}=1.6108$, $T=0.032551$ for $(0.3,0.9)$. The small branch is not. Comparing the free energy $F=M-TS$ at equal temperature, the small branch lies above the stable charged branch by $0.197$ at $T=0.01$ for $(0.9,0.3)$ and by $0.027$ for the dual point. Both small branches are thermodynamically disfavored and locally unstable, but their separation from the stable branch differs by a factor of seven under the exchange of the charges.

\section{Geodesic structure}\label{isec5}

Test particles probe the geometry at the radii where they move, so they separate cleanly the two regimes found above. Circular and bound orbits live at a few $M$, where Eq.~\eqref{eq:Deltaf} predicts a tiny odd part; radially infalling particles reach the core, where the asymmetry is of order one. This section follows both, from the photon sphere through precessing and zoom-whirl orbits to radial infall.

\subsection{Photon sphere, Lyapunov exponent and ISCO}\label{isec5a}

For the metric~\eqref{eq:metric}, circular null orbits satisfy $rf'=2f$ \cite{Claudel:2000yi}, which with $f=1-2m/r$ becomes
\begin{equation}
r_{\rm ph}\big[1+m'(r_{\rm ph})\big]=3\,m(r_{\rm ph}),\qquad b_c=\frac{r_{\rm ph}}{\sqrt{f(r_{\rm ph})}} .\label{eq:rph}
\end{equation}
The instability of the orbit is measured by the Lyapunov exponent \cite{Cardoso:2008bp}
\begin{equation}
\lambda=\sqrt{\frac{f(r_{\rm ph})\,\big[2f(r_{\rm ph})-r_{\rm ph}^2f''(r_{\rm ph})\big]}{2r_{\rm ph}^2}} ,\label{eq:lyap}
\end{equation}
which governs the eikonal damping of quasinormal modes. For massive particles the specific energy and angular momentum of a circular orbit are
\begin{equation}
\mathcal{E}^2=\frac{2f^2}{2f-rf'},\qquad \mathcal{L}^2=\frac{r^3f'}{2f-rf'},\label{eq:EL}
\end{equation}
and the ISCO solves
\begin{equation}
r f f''-2rf'^2+3ff'=0 .\label{eq:isco}
\end{equation}
We evaluate $f'$ and $f''$ from the exact identities $f'=2m/r^2-2m'/r$ and $f''=-4m/r^3+4m'/r^2-2m''/r$, which need only $m$ from either route and the elementary $m'$.

To first order in the coupling the photon sphere and the critical impact parameter follow in closed form. With $q^2\equiv Q^2+P^2$, the Reissner-Nordstr\"om values are $r_0=\frac12\big(3M+\sqrt{9M^2-8q^2}\big)$, $f_0=1-2M/r_0+q^2/r_0^2$ and $b_0=r_0/\sqrt{f_0}$. Treating the $r^{-6}$ term of Eq.~\eqref{eq:fseries} as a perturbation gives
\begin{equation}
r_{\rm ph}\simeq r_0+\frac{16\,\mathcal{K}}{5r_0^3\,(6Mr_0-8q^2)},\qquad b_c\simeq b_0+\frac{\mathcal{K}}{5\,r_0^5\,f_0^{3/2}} .\label{eq:bcfirst}
\end{equation}
Both corrections depend on the charges only through $q^2$ and $\mathcal{K}$ and are therefore duality-even. For $Q=P=0.4M$ and $a=0.05M^2$, Eq.~\eqref{eq:bcfirst} gives $r_{\rm ph}=2.768954M$ and $b_c=4.8991972M$, against the exact $2.768954M$ and $4.8991971M$; at $a=0.1M^2$ and $Q=P=0.6M$ the error in $b_c$ grows to $2\times10^{-5}M$, the size of the neglected second-order terms.

Table~\ref{tab2} lists the results for Schwarzschild, for the two single-charge configurations of equal magnitude, and for two dual pairs. The single-charge pair $(0.4,0)M$ and $(0,0.4)M$ agrees to eight digits in $b_c$, $\lambda$ and $r_{\rm ISCO}$, and the dual pair $(0.7,0.3)M$, $(0.3,0.7)M$ at $a=0.1M^2$ differs in $b_c$ by $1.1\times10^{-6}M$ and in $\lambda$ by $9\times10^{-7}/M$.

\begin{figure}[t]
\centering
\includegraphics[width=\columnwidth]{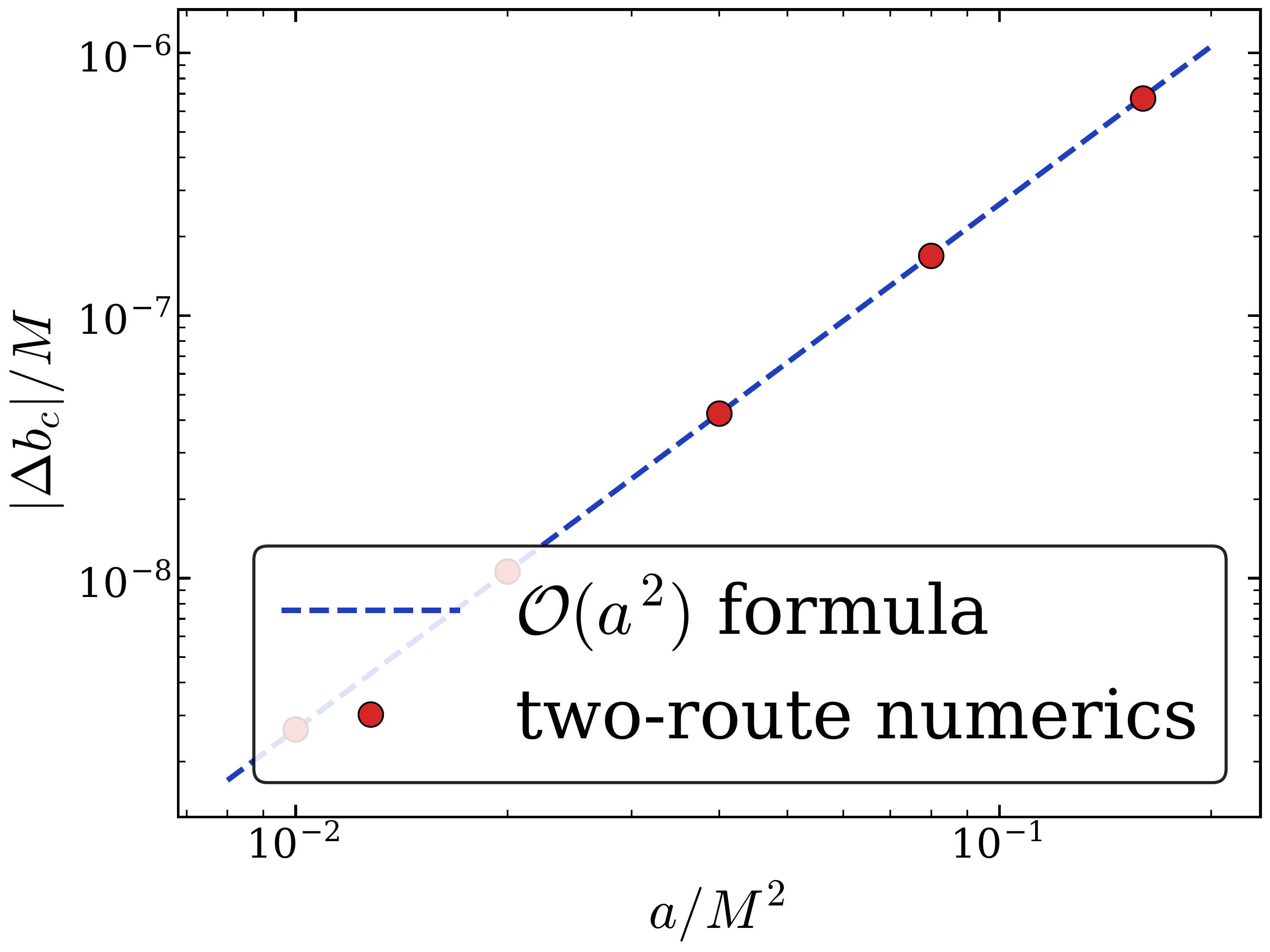}
\caption{Duality-odd shift of the critical impact parameter, $|b_c(Q,P)-b_c(P,Q)|$, for $(Q,P)=(0.6,0.2)M$. Points are the difference of two route-II evaluations at 30-digit precision; the dashed line is Eq.~\eqref{eq:dbc} with the second-order odd term of Eq.~\eqref{eq:DeltafEEH}. The slope is two on the logarithmic axes.}
\label{fig5}
\end{figure}

The size of the odd shift follows from Eq.~\eqref{eq:Deltaf}. Because $b_c^2=\min(r^2/f)$, a small change $\delta f$ shifts the critical impact parameter, to first order and without moving $r_{\rm ph}$, by
\begin{equation}
\delta b_c=-\frac{b_c\,\delta f(r_{\rm ph})}{2f(r_{\rm ph})} .\label{eq:dbc}
\end{equation}
With $\delta f$ given by Eq.~\eqref{eq:DeltafEEH} this predicts $\Delta b_c\equiv b_c(Q,P)-b_c(P,Q)\propto a^2$. Figure~\ref{fig5} compares the prediction with route II for $(Q,P)=(0.6,0.2)M$ over $0.01\leq a/M^2\leq0.16$. The ratio is $0.99932$ at the smallest coupling and $0.98925$ at the largest; the departure is the next-order term of Eq.~\eqref{eq:fseries}.

\begin{table*}[t]
\caption{Exterior observables for $M=1$ and $b=7a/4$. Horizon radius (routes I and II agree in every digit shown), Hawking temperature, photon-sphere radius, critical impact parameter, Lyapunov exponent, ISCO radius and the specific energy and angular momentum at the ISCO. Members of a pair related by $Q\leftrightarrow P$ differ only in the last digits shown. For $Q=0$ the closed form~\eqref{eq:fmag} replaces route II.}
\label{tab2}
\begin{ruledtabular}
\small
\setlength{\tabcolsep}{4pt}
\begin{tabular}{ccccccccccc}
\textbf{$Q$} & \textbf{$P$} & \textbf{$a$} & \textbf{$r_{\rm h}$} & \textbf{$T$} & \textbf{$r_{\rm ph}$} & \textbf{$b_c$} & \textbf{$\lambda$} & \textbf{$r_{\rm ISCO}$} & \textbf{$\mathcal{E}_{\rm ISCO}$} & \textbf{$\mathcal{L}_{\rm ISCO}$}\\
\hline
0 & 0 & 0.05 & 2.00000000 & 0.0397887 & 3.00000000 & 5.19615242 & 0.1924501 & 6.0000000 & 0.9428090 & 3.4641016\\
0.4 & 0 & 0.05 & 1.91653579 & 0.0397150 & 2.88925497 & 5.05298409 & 0.1940744 & 5.7527863 & 0.9405700 & 3.3847215\\
0 & 0.4 & 0.05 & 1.91653584 & 0.0397150 & 2.88925498 & 5.05298409 & 0.1940744 & 5.7527863 & 0.9405700 & 3.3847215\\
0.4 & 0.4 & 0.05 & 1.82481599 & 0.0394388 & 2.76895370 & 4.89919709 & 0.1954273 & 5.4889524 & 0.9379888 & 3.2998714\\
0.7 & 0.3 & 0.1 & 1.64996406 & 0.0381667 & 2.54487939 & 4.61869938 & 0.1963278 & 5.0125161 & 0.9327526 & 3.1466931\\
0.3 & 0.7 & 0.1 & 1.64998597 & 0.0381709 & 2.54488209 & 4.61870049 & 0.1963287 & 5.0125163 & 0.9327526 & 3.1466932\\
0.6 & 0.6 & 0.1 & 1.53500958 & 0.0366621 & 2.40238787 & 4.44536607 & 0.1953586 & 4.7214126 & 0.9291335 & 3.0534999\\
\end{tabular}
\end{ruledtabular}
\end{table*}

\subsection{Bound orbits, periastron precession and zoom-whirl motion}\label{isec5b}

For a bound timelike orbit with periastron $r_p$ and apastron $r_a$, the turning-point conditions $\mathcal{E}^2=f(r)\,(1+\mathcal{L}^2/r^2)$ at both radii fix
\begin{equation}
\mathcal{L}^2=\frac{f(r_a)-f(r_p)}{f(r_p)/r_p^2-f(r_a)/r_a^2},\qquad \mathcal{E}^2=f(r_p)\Big(1+\frac{\mathcal{L}^2}{r_p^2}\Big),\label{eq:ELbound}
\end{equation}
and the advance of the periastron per radial period is
\begin{equation}
\Delta\varphi=2\int_{r_p}^{r_a}\frac{\mathcal{L}\,dr}{r^2\sqrt{\mathcal{E}^2-f(r)\,(1+\mathcal{L}^2/r^2)}}-2\pi .\label{eq:precession}
\end{equation}
We evaluate Eq.~\eqref{eq:precession} after the substitution $r=r_p+(r_a-r_p)\sin^2\chi$, which removes the endpoint singularities, and check it against direct integration of the orbit equation
\begin{equation}
\frac{d^2u}{d\varphi^2}=-\frac12\frac{d}{du}\Big[F(u)\Big(\frac{1}{\mathcal{L}^2}+u^2\Big)\Big],\label{eq:orbiteq}
\end{equation}
with $u=1/r$ and $F(u)=f(1/u)$,
and locating successive apastra. The two routes agree to $3\times10^{-12}$ rad.

\begin{figure*}[t]
\centering
\includegraphics[width=\textwidth]{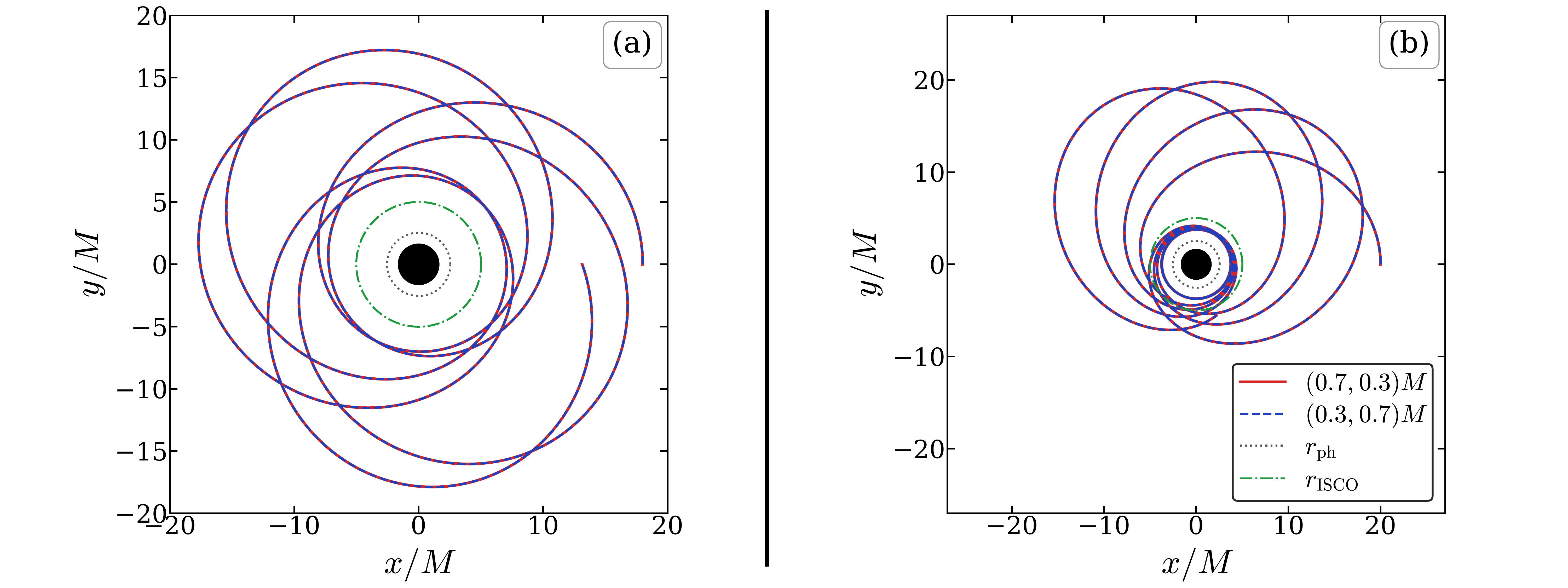}
\caption{Timelike orbits around the dyon $(Q,P)=(0.7,0.3)M$ (solid) and its dual $(0.3,0.7)M$ (dashed) for $a=0.1M^2$. (a) Precessing rosette with $r_p=7M$, $r_a=18M$, over six radial periods. (b) Zoom-whirl orbit with $r_p=3.72M$, $r_a=20M$, over three radial periods. The black disk is the event horizon, the dotted circle the photon sphere and the dash-dotted circle the ISCO. In both panels the two orbits coincide to plotting accuracy, as the precession values of Table~\ref{tab3} require.}
\label{fig6}
\end{figure*}

Table~\ref{tab3} compares Schwarzschild, Reissner-Nordstr\"om with the same $Q^2+P^2$, and the dual EEH pair at $a=0.1M^2$. The Euler-Heisenberg term changes the precession relative to Reissner-Nordstr\"om by $2\times10^{-4}$ to $3\times10^{-6}$ rad, growing toward the hole as $r_p^{-6}$ would suggest. The exchange of the charges changes it by $3\times10^{-8}$ to $4\times10^{-11}$ rad, four orders of magnitude less, in line with the $r^{-10}$ falloff of the odd term. Figure~\ref{fig6}(a) shows a rosette orbit with $r_p=7M$ and $r_a=18M$; the two members of the dual pair are drawn on top of each other and cannot be separated by eye after six radial periods. Close to the unstable circular orbit the motion turns into zoom-whirl behavior, studied for other quantum-corrected and dressed black holes in Ref.~\cite{AlBadawi:2026zoom}. For $r_p=3.72M$ and $r_a=20M$ the orbit completes more than two whirls near the photon sphere before zooming out, and Fig.~\ref{fig6}(b) shows that even this strongly relativistic motion is duality-blind.

\begin{table*}[t]
\caption{Periastron advance $\Delta\varphi$ (rad) per radial period for $M=1$ and $a=0.1$, from Eq.~\eqref{eq:precession}. Reissner-Nordstr\"om (RN) uses $Q^2+P^2=0.58$, the value shared by the dual EEH pair.}
\label{tab3}
\begin{ruledtabular}
\begin{tabular}{ccccc}
\textbf{$(r_p,r_a)$} & \textbf{Schwarzschild} & \textbf{RN} & \textbf{EEH $(0.7,0.3)$} & \textbf{EEH $(0.3,0.7)$}\\
\hline
$(5.5,14)$ & 7.111487884 & 4.524551986 & 4.524770961 & 4.524770993\\
$(7,18)$ & 3.681199831 & 2.799504307 & 2.799540427 & 2.799540429\\
$(10,30)$ & 1.847276656 & 1.534196276 & 1.534199568 & 1.534199568\\
\end{tabular}
\end{ruledtabular}
\end{table*}

\subsection{Radial infall toward the core}\label{isec5c}

Radial timelike geodesics reach the region where the duality-odd structure is of order one. Since $\partial_t$ is a Killing vector everywhere, the conserved energy $\mathcal{E}$ obeys
\begin{equation}
\Big(\frac{dr}{d\tau}\Big)^2=\mathcal{E}^2-f(r)\label{eq:radial}
\end{equation}
in every region, including those between horizons where $t$ is spacelike. A particle released from rest at $r_0$ has $\mathcal{E}^2=f(r_0)$ and turns around wherever $f(r)=\mathcal{E}^2$ inside. Whether it bounces or reaches $r=0$ is therefore read off from $f$ alone, and the proper time to the turning point is
\begin{equation}
\tau(r)=\int_{r}^{r_0}\frac{ds}{\sqrt{\mathcal{E}^2-f(s)}} .\label{eq:tau}
\end{equation}

\begin{figure*}[t]
\centering
\includegraphics[width=\textwidth]{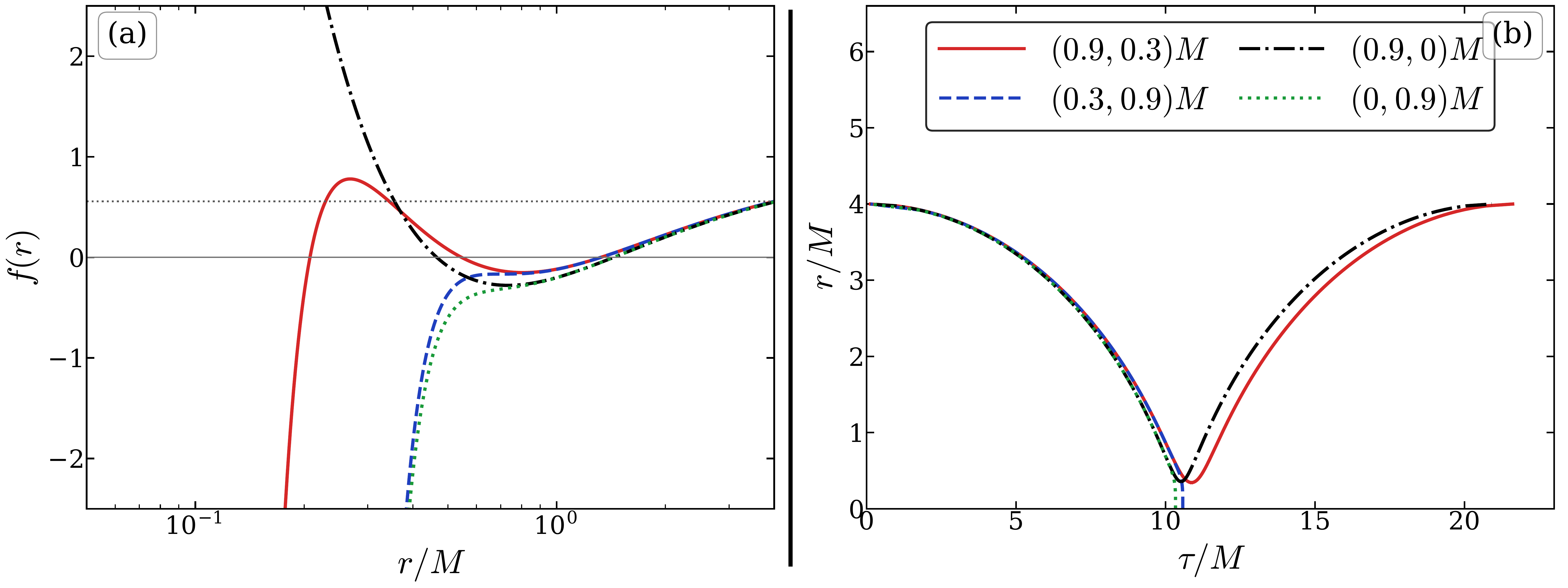}
\caption{Radial infall from rest at $r_0=4M$ for $a=0.05M^2$. (a) Metric function on a logarithmic radial axis, with the dotted line at $\mathcal{E}^2=f(r_0)$ for the dyons. A particle turns around where the curve crosses the line from below. (b) Radius against proper time. The electric configuration and the electrically dominated dyon bounce at $r\simeq0.35M$ and return outward through the horizons, while the magnetic configuration and the magnetically dominated dyon reach $r=0$.}
\label{fig7}
\end{figure*}

Figure~\ref{fig7} shows the result for release at $r_0=4M$ and $a=0.05M^2$. For the electric configuration $(0.9,0)M$, whose metric function rises to $+\infty$ at the center, the particle bounces at $r=0.3567M$ after $\tau=10.52M$, as in Reissner-Nordstr\"om. For the magnetic configuration $(0,0.9)M$ it reaches $r=0$ after $\tau=10.34M$. The dual dyons behave like their dominant charge. The dyon $(0.9,0.3)M$ has three horizons at $0.2076M$, $0.5452M$ and $1.3260M$, and between the inner two the metric function rises above $\mathcal{E}^2=0.5562$, so the particle bounces at $r=0.3441M$ after $\tau=10.87M$ without meeting the magnetically supported horizon. Its dual $(0.3,0.9)M$ has a single horizon and the particle reaches the center after $\tau=10.58M$. The fate of an infalling particle is thus decided by the orientation of the charge at order one, even though the two dyons are indistinguishable from outside. The comparison is the geodesic counterpart of the horizon count of Fig.~\ref{fig2}(b), and it parallels the charge-dependent fall of test bodies discussed in Ref.~\cite{Halilsoy:2024plate}.

\section{Lensing and the shadow}\label{isec6}

This section turns the metric expansion into a weak-deflection formula with its odd part, confronts the shadow with the Event Horizon Telescope image of Sagittarius A*, and identifies the one optical quantity that does respond to the exchange of the charges.

\subsection{Weak deflection}\label{isec6a}

A perturbation $\delta f=A_n r^{-n}$ of the metric function changes the deflection angle at leading order by
\begin{equation}
\delta\hat\alpha_n=-A_n\,\frac{\sqrt\pi\,\Gamma\big(\frac{n+3}{2}\big)}{\Gamma\big(\frac n2+1\big)}\,\frac{1}{\beta^n},\label{eq:alphan}
\end{equation}
with $\beta$ the impact parameter. For $n=1$ and $n=2$ this reproduces the Schwarzschild term $4M/\beta$ and the charge term $-3\pi Q^2/(4\beta^2)$; we checked the cases $n=6$ and $n=10$ against direct quadrature of the orbit integral to four significant figures. Applied to Eq.~\eqref{eq:fseries},
\begin{align}
\hat\alpha={}&\frac{4M}{\beta}+\frac{\pi\big[15M^2-3(Q^2+P^2)\big]}{4\beta^2}\nonumber\\
&+\frac{7\pi\mathcal{K}}{16\beta^6}-\frac{77\pi\,Q^2X^2}{32\,\beta^{10}}+\cdots,\label{eq:alpha}
\end{align}
where the dots denote terms of higher order in $M/\beta$ and in the charges. The duality-odd part of the deflection is
\begin{align}
\hat\alpha(Q,P)-\hat\alpha(P,Q)={}&-\frac{77\pi a^2}{128\,\beta^{10}}\,(Q^2-P^2)\nonumber\\
&\times(4Q^4-Q^2P^2+4P^4)\label{eq:dalpha}
\end{align}
for EEH. It is suppressed by $(a/\beta^2)^2$ relative to the $\mathcal{K}$ term, so weak lensing measures $Q^2+P^2$ and $\mathcal{K}$ and nothing else. The optical Gauss-Bonnet method \cite{Gibbons:2008rj} offers a third route to these coefficients that we have not pursued here. In the strong-deflection limit \cite{Virbhadra:1999nm,Bozza:2002zj} the coefficients are fixed by $r_{\rm ph}$ and $b_c$ and inherit the duality blindness of Table~\ref{tab2}.

\subsection{Shadow and the Event Horizon Telescope}\label{isec6b}

\begin{figure}[t]
\centering
\includegraphics[width=\columnwidth]{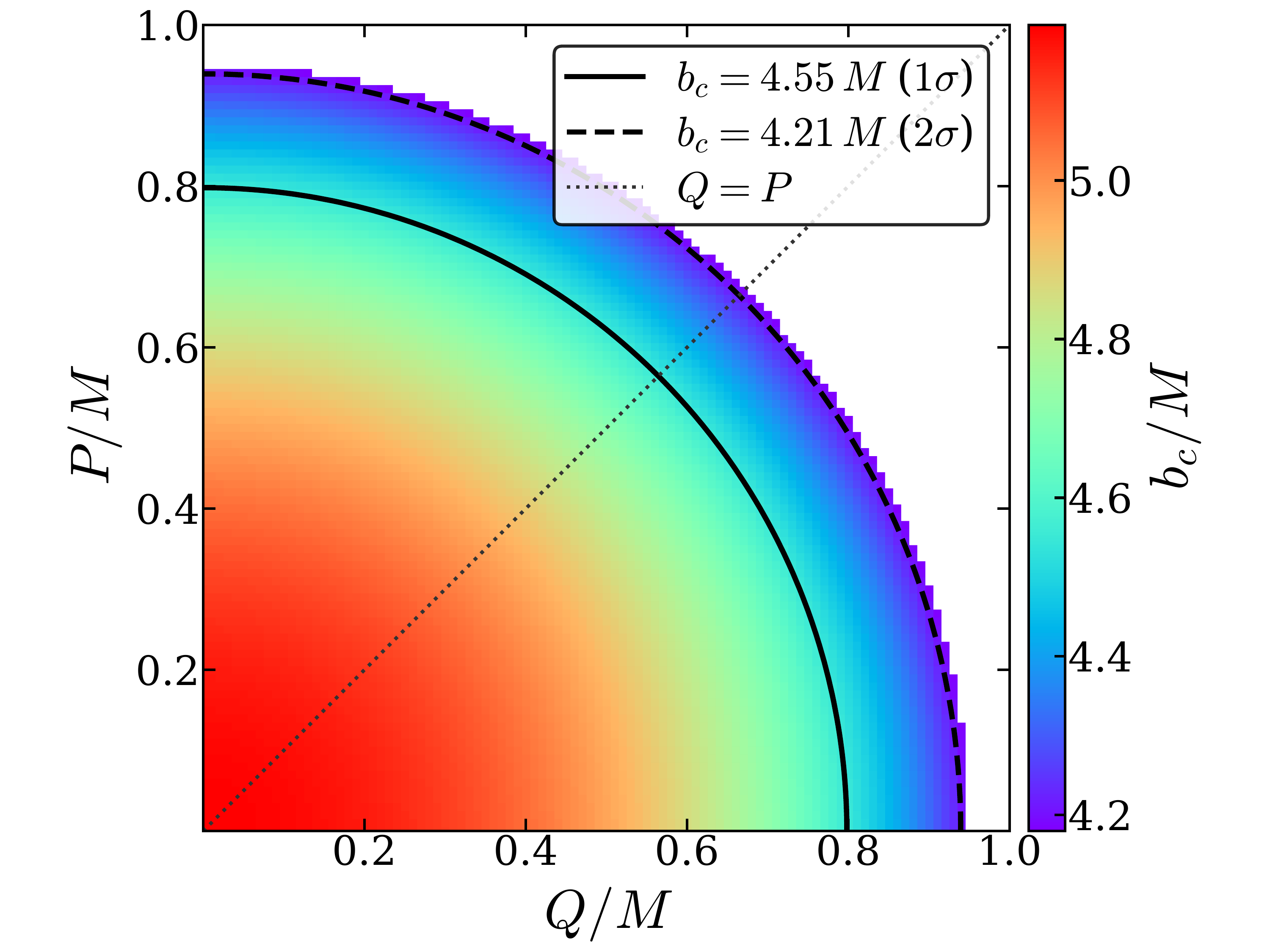}
\caption{Critical impact parameter over the charge plane for $a=0.05M^2$ and $b=7a/4$, restricted to $\sqrt{Q^2+P^2}\leq0.95M$. The solid and dashed contours are the $1\sigma$ and $2\sigma$ lower bounds from the Event Horizon Telescope image of Sagittarius A*. The map is symmetric about the dotted diagonal to better than $10^{-5}M$.}
\label{fig8}
\end{figure}

For an asymptotically flat, static black hole and a distant observer the shadow radius equals $b_c$ \cite{Perlick:2021aok}. The Event Horizon Telescope image of Sagittarius A* \cite{EventHorizonTelescope:2022wkp,EventHorizonTelescope:2022xqj}, combined with the Keck and VLTI mass-to-distance ratios, gives $4.55\leq b_c/M\leq5.22$ at $1\sigma$ and $4.21\leq b_c/M\leq5.56$ at $2\sigma$ \cite{Vagnozzi:2022moj}, in line with earlier shadow-size tests \cite{Johannsen:2015mdd,Liu:2026univ,Sakalli:2026dunkl}. Figure~\ref{fig8} shows $b_c$ over the charge plane for $a=0.05M^2$. The level curves are circles to the eye, and the bounds become
\begin{equation}
\sqrt{Q^2+P^2}<\begin{cases}0.798M & (1\sigma),\\ 0.939M & (2\sigma).\end{cases}\label{eq:EHT}
\end{equation}
The bounds on the electric and magnetic axes agree to five digits; the small difference along the diagonal, below $3\times10^{-4}M$, comes from the even quartic $\mathcal{K}$. The bounds are close to the Reissner-Nordstr\"om ones \cite{Ghosh:2023kge,Tsukamoto:2024kase,Mishra:2024rn}, as expected from the smallness of $\mathcal{K}/r_{\rm ph}^6$. A shadow measurement therefore constrains the total charge but cannot tell an electric black hole from a magnetic one of the same magnitude, in line with earlier magnetic EEH bounds \cite{Allahyari:2019jqz}.

\subsection{Capture beyond the charged branch}\label{isec6c}

\begin{figure*}[t]
\centering
\includegraphics[width=\textwidth]{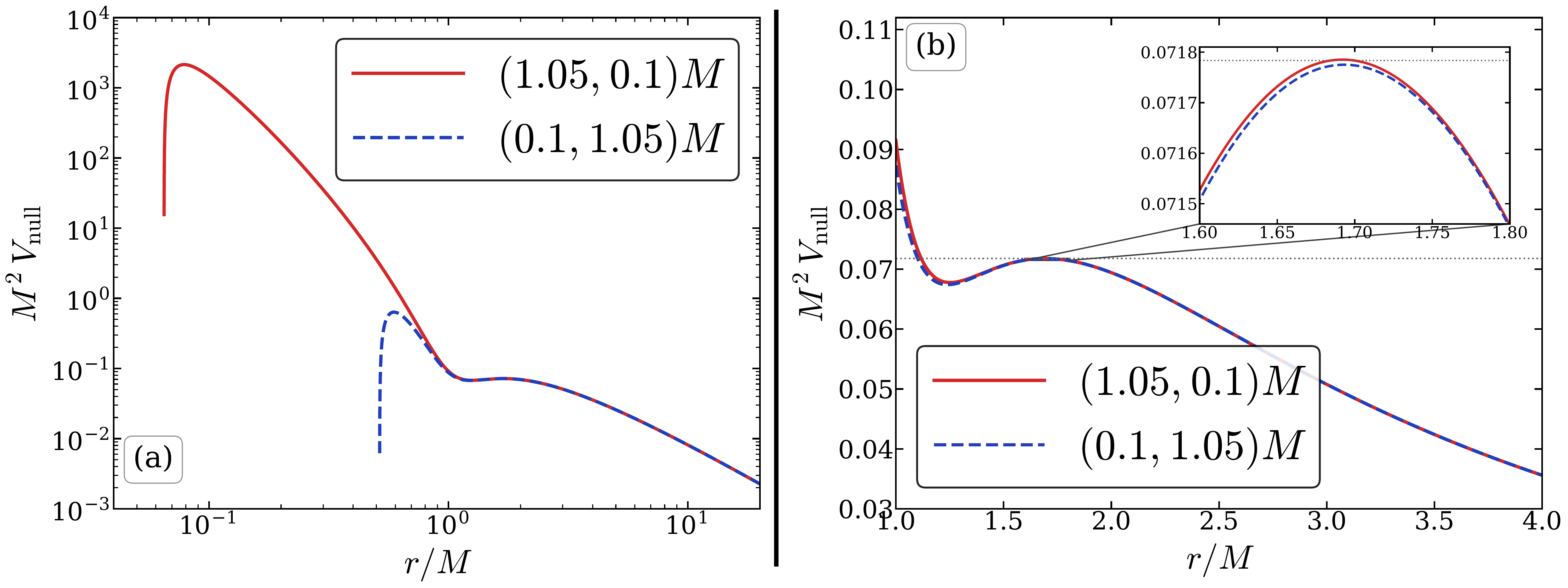}
\caption{Null effective potential $V=f/r^2$ beyond the charged branch, $a=0.05M^2$, for $(Q,P)=(1.05,0.1)M$ (solid) and $(0.1,1.05)M$ (dashed), each plotted from its event horizon outward. (a) Logarithmic axes: the inner barrier of the electrically dominated configuration is three orders of magnitude higher than that of its dual. (b) Linear axes near the photon sphere at $r=1.692M$, with the dotted line at $1/b_c^2$. The inset magnifies the peak, where the two curves separate by about $10^{-5}/M^2$.}
\label{fig9}
\end{figure*}

The duality blindness of the critical curve rests on the outer horizon being the large charged one. Past the three-horizon window, where the event horizon has jumped inward, the effective potential $V=f/r^2$ develops an inner barrier near the small horizon in addition to the outer photon sphere. For $(Q,P)=(1.05,0.1)M$ the outer unstable orbit sits at $r=1.692M$ with $b=3.7324M$, and the inner barrier peaks at $r=0.079M$ with $b=0.0216M$. For the dual $(0.1,1.05)M$ the outer orbit agrees to within $3\times10^{-4}M$, with $b=3.7326M$, but the inner barrier sits at $r=0.592M$ with $b=1.255M$. The photon ring is still duality-blind, while the capture cross section $\pi b_{\rm in}^2$, which sets the size of the dark core, differs by more than three orders of magnitude. Figure~\ref{fig9} shows the two potentials. On a logarithmic scale the inner barrier of the electrically dominated configuration rises to $M^2V\simeq2\times10^3$, while that of its dual barely exceeds the outer peak; on a linear scale near the photon sphere the two curves are indistinguishable.

\begin{figure*}[t]
\centering
\includegraphics[width=\textwidth]{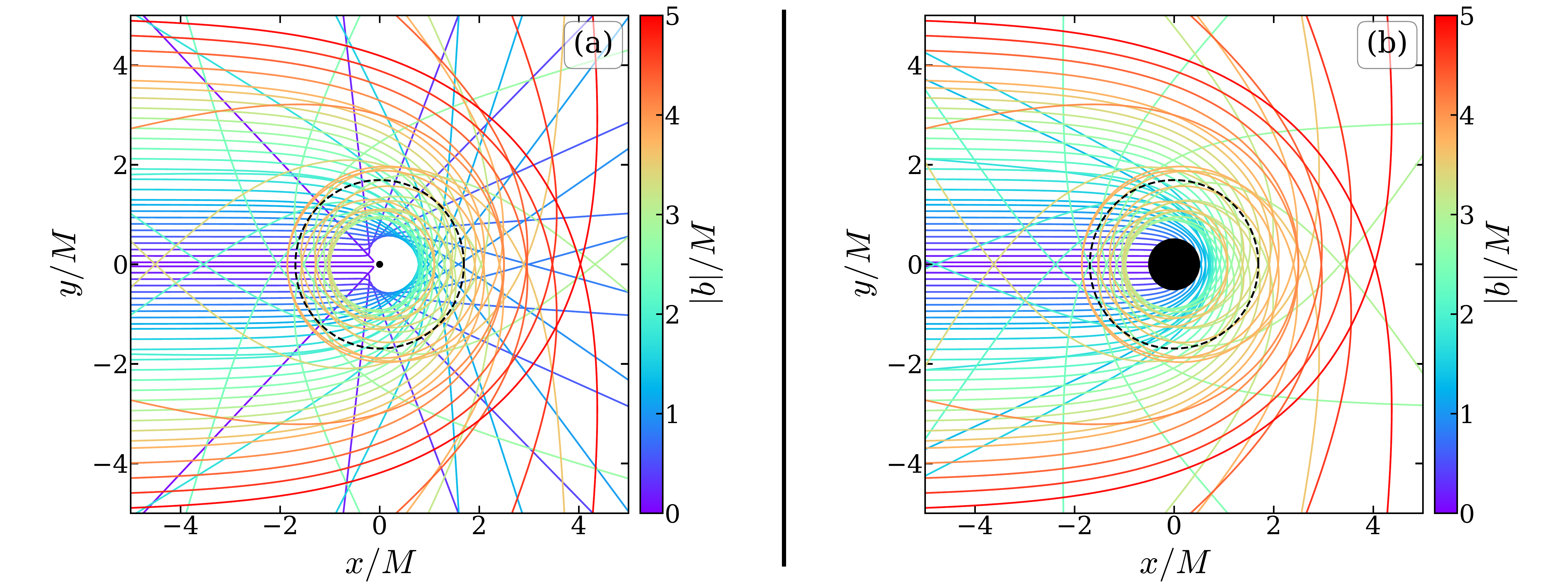}
\caption{Null geodesics arriving from the left with impact parameter $|b|\leq5M$, colored by $|b|$, for $a=0.05M^2$. (a) $(Q,P)=(1.05,0.1)M$: nearly all rays reflect off the inner barrier around the tiny horizon (black dot). (b) $(Q,P)=(0.1,1.05)M$: rays with $|b|<1.255M$ fall onto the horizon (black disk). The dashed circle marks the common outer photon sphere.}
\label{fig10}
\end{figure*}

Figure~\ref{fig10} traces the null geodesics that fall onto the two configurations from the left, with impact parameters up to $5M$. For the magnetically dominated dyon every ray with $|b|<1.255M$ ends on the horizon of radius $0.516M$, rays between the inner and outer barriers wind around the photon sphere and reflect, and rays beyond $3.73M$ are deflected. For the electrically dominated dyon almost every ray that crosses the photon sphere reflects off the inner barrier and returns to infinity, since only $|b|<0.022M$ reaches the horizon of radius $0.065M$. An observer would see the same photon ring in both cases and a dark core that is either about the size of the horizon or nearly absent. These configurations lie outside the bounds~\eqref{eq:EHT} and in the strong-field region where the quadratic model may fail, as discussed in Sec.~\ref{isec8}. We report them to show that the core asymmetry has an optical counterpart once the outer horizon is gone.

\section{Conclusion}\label{isec8}

We used the exact dyonic solution of the Einstein-Euler-Heisenberg equations to locate where the breaking of electric-magnetic duality lives, and the answer splits cleanly along the radius. Outside the photon sphere the black hole is nearly blind to the orientation of its charge. The first-order metric depends on the charges only through $Q^2+P^2$ and the symmetric quartic $\mathcal{K}$; the odd part begins at second order in the coupling with a tenth-power radial falloff, and we derived its coefficient for the whole quadratic family. The shifts it produces in the shadow, the Lyapunov exponent, the ISCO and the periastron advance stay at the level of $10^{-6}$ in units of $M$, or below, for couplings as large as a tenth of $M^2$. Precessing and zoom-whirl orbits of a dyon and its dual coincide to plotting accuracy, weak lensing measures only $Q^2+P^2$ and $\mathcal{K}$, and the Event Horizon Telescope image of Sagittarius A* bounds the total charge, $\sqrt{Q^2+P^2}<0.798M$ at $1\sigma$ for $a=0.05M^2$, while saying nothing about its orientation.

Near the center, the asymmetry is of order one. The magnetic field is fixed by topology and is never screened, while the electric field is screened by the cubic Gauss constraint. Any magnetic charge therefore forces the metric function to $-\infty$ at the origin, guarantees an odd number of horizons and rules out a naked singularity, whereas an electric configuration lighter than its field energy, which we found in closed form, has an even number of horizons, two or none, and can be horizonless. Whether an extremal limit exists is decided by a single function of the charges, the supremum of $m'(r)$, whose critical curve passes through the origin on the electric side and crosses the magnetic axis at $3\sqrt{3a/2}$. The same function controls the three-horizon window, the inward jump of the event horizon at its edge, and a small-black-hole branch of negative heat capacity that exists only for $P\neq0$. Radial infall makes the asymmetry concrete: a particle dropped onto the electrically dominated dyon bounces inside the static pocket between its inner horizons, while on the dual dyon it reaches the center. Past the charged branch the same core asymmetry appears in the capture cross section, which differs by more than three orders of magnitude between a dyon and its dual, although the photon ring stays the same. The first law and the Smarr relation hold with closed-form potentials once the coupling is treated as a thermodynamic variable.

Each headline number rests on at least two independent evaluations. The radial quadrature and the $z$ representation of the Lauricella solution agree to $10^{-14}$ in the metric function, a third evaluation in Wolfram Language reproduces the horizon, photon-sphere and shadow radii, the odd coefficient of Eq.~\eqref{eq:Deltaf} matches the numerical difference to $2\times10^{-7}$ at $r=40M$, the precession integral matches the integrated orbit to $3\times10^{-12}$ rad, and the background reproduces Schwarzschild, Reissner-Nordstr\"om, the exact magnetic solution and the integrable locus of Ref.~\cite{Ahmed:2026ufj}. What is universal and what is model dependent can be stated precisely. For every Lagrangian of the quadratic family with $a>0$ and $c\geq0$ the first-order metric is duality-even, the odd part starts as $a^2r^{-10}$ with the coefficient of Eq.~\eqref{eq:Deltaf}, and any magnetic charge forces $f\to-2aP^4/(5r^6)$ at the center, independently of $Q$ and $b$. The location of the critical curve, the widths of the three-horizon windows and the size of every shift depend on $b/a$ and on the chosen charges and coupling.

The scope of these statements should be kept in view. The coupling $a$ is treated as a free length scale; for quantum electrodynamics it is fixed by the fine-structure constant and the electron mass and is many orders of magnitude below $M^2$ for astrophysical black holes, so our values were chosen to make the structure visible, and only the universal statements carry over unchanged. The Euler-Heisenberg Lagrangian is the weak-field limit of the one-loop effective action \cite{Dunne:2004nc}, so results about the region $r\lesssim(aP^2)^{1/4}$, including the inner horizons, the bounce of infalling particles and the inner capture barrier, describe the classical quadratic model rather than quantum electrodynamics. We traced null geodesics of the background metric; photons of the nonlinear field itself follow the effective metric of Ref.~\cite{Novello:1999pg}, whose birefringent corrections enter at the same order as $\mathcal{K}$. The conditions $a>0$ and $c>0$ guarantee a single real root of the Gauss constraint, the causality and energy conditions of Ref.~\cite{Russo:2024xnh} hold in the weak-field domain, and our treatment covers static configurations without a perturbative stability analysis.

In the near future, we plan to compute the birefringent photon propagation on the effective metric of the nonlinear field, which enters the shadow at the same order as $\mathcal{K}$ and may carry its own odd part, and to obtain the quasinormal spectrum of scalar and electromagnetic perturbations on the exact background, testing whether the eikonal correspondence with the Lyapunov exponent survives near the magnetically supported horizon. We also intend to extend the extremality functional to slowly rotating dyons and to the anti-de Sitter family, where the cusp of Eq.~\eqref{eq:Pc} should meet the critical point of the extended phase space, and to follow charged test particles, whose Lorentz force will couple directly to the two charges and may expose the asymmetry outside the core.

\section*{Acknowledgments}
\.{I}.S. is grateful to Eastern Mediterranean University (EMU), the Scientific and Technological Research Council of T\"urkiye (T\"UB\.{I}TAK), the Anatolian University Libraries Consortium (ANKOS), and the Sponsoring Consortium for Open Access Publishing in Particle Physics (SCOAP3) for their support. He also acknowledges the networking support of COST Actions CA18108, CA22113, CA21106, CA23130, CA23115, and CA21136.

\section*{Data availability}
No observational data were generated in this study. The computational worksheets used for the numerical evaluations, figures and tables are available from the corresponding author upon reasonable request.

\bibliographystyle{apsrev4-2}
\bibliography{ref}

\end{document}